# New Results from RENO using 1500 Days of Data

Seon-Hee Seo[1, a)] for the RENO Collaboration

[1] Department of Physics & Astronomy, Seoul National University, 1 Gwanak-ro, Gwanak-gu, Seoul 08826, Korea

[a)] Corresponding author: sunny.seo@snu.ac.kr

**Abstract**. RENO (Reactor Experiment for Neutrino Oscillation) is the first reactor neutrino experiment which began data-taking in 2011 with two identical near and far detectors in Yonggwang, Korea. Using 1500 live days of data, $\sin^2 2\theta_{13}$ and $|\Delta m^2_{ee}|$ are updated using spectral measurements: $\sin^2 2\theta_{13} = 0.086 \pm 0.006$ (stat.) $\pm 0.005$ (syst.) and $|\Delta m^2_{ee}| = 2.61^{+0.15}_{-0.16}$ (stat.) $\pm 0.09$ (syst.) (x$10^{-3}$ eV$^2$). The correlation between the 5 MeV excess rate and the reactor thermal power is again clearly observed with the increased data set.

## 1. Introduction

Neutrino oscillation is well known and proven. All three neutrino mixing angles and two mass square differences in the PMNS matrix are measured. However, precise measurements of these parameters are still important for current and future neutrino oscillation experiments to measure leptonic CP violation and to determine neutrino mass ordering. In this work RENO improved background systematic uncertainty and updated $\sin^2 2\theta_{13}$ and $|\Delta m^2_{ee}|$ using 1500 live days of data collected in the detectors. The study of the 5 MeV excess is also updated.

## 2. RENO Experimental Setup and Data Set

RENO is located in the southwest region of South Korea where there is a commercial nuclear reactor power plant consisting of six reactor cores which produce total 16.8 GW thermal power in full operation mode. RENO near (far) detector is built under a small hill with 120 (450) m.w.e. overburden that has 290 (1380) m baseline from the center of the linear array of the six reactors. The two detectors are built identically in a cylindrical shape. Each detector consists of inner detector and outer veto detector (350 ton purified water). The inner detector is classified as target (16 ton liquid scintillator with 0.1 % Gd), γ-catcher (29 ton liquid scintillator), and buffer (65 ton mineral oil) where total 354 Hamamatsu 10 inch PMTs are mounted. More details on the RENO experimental setup and the detector are found in [1].

RENO has been taking data since August 2011 continuously with average DAQ live time efficiency of ~95% for both detectors. This analysis uses data collected from Aug. 19, 2011 to Apr. 23, 2017 for near detector and from Aug. 11, 2011 to Sept. 23, 2015 for far detector. Total live time of the data is 1547.35 (1397.78) days for near (far) detector.

## 3. Event Selection and Remaining Background

A reactor neutrino arriving at detector can interact with a free proton in target or γ-catcher and then produces a positron and a neutron if the neutrino energy is greater than 1.8 MeV. In this Inverse Beta Decay (IBD) process a positron annihilates immediately and produces a prompt signal while a neutron needs to be thermalized until it is captured by either Gd or Hydrogen and produces a delayed signal.

The average delay time is about 30 (180) μs for n-Gd (n-H). In this analysis only IBD n-Gd events are selected to get cleaner sample of signal neutrino events among accidental, fast neutron, $^9$Li/$^8$He backgrounds. $^{252}$Cf contamination was accidentally introduced in our detector while taking source data for energy calibration during October 2012 and this gives us additional background since $^{252}$Cf produces multiple neutrons as well as spontaneous γ-rays.

*3.1. Event selection*
Events are selected by applying the IBD selection criteria described in [2]. In this analysis, to reduce background rate and its uncertainty, we slightly optimized the values of the spatial coincidence requirement of ΔR < 2.0 m to reduce the accidental background. The following multiplicity requirements are also changed to make additional reduction of fast neutron, $^9$Li/$^8$He and $^{252}$Cf backgrounds (note that the indexes of changed criteria are the ones used in [2]): a timing veto requirement for rejecting coincidence pair (a) if they are accompanied by any preceding ID or OD trigger within a 300 μs window before their prompt candidate, (b) if they are followed by any subsequent ID-only trigger other than those associated with the delayed candidate within a 200 (800) μs window from their prompt candidate (only far $^{252}$Cf contaminated data), (d) if there are other subsequent pairs within the 500 (1,000) μs interval (only far $^{252}$Cf contaminated data), (f) if they are accompanied by a prompt candidate of $E_P$ > 3 MeV and $Q_{max}/Q_{tot}$ < 0.04 within a 10 (20) s window and a distance of 40 (50) cm for near (far) $^{252}$Cf contaminated data; (ix) a spatial veto requirement for rejecting coincidence pairs in the far detector only if the vertices of their prompt candidates are located in a cylindrical volume of 50 cm in radius, centered at x = +12.5 cm and y = +12.5 cm and z < -110 cm. Total dead time due to the selection criteria is estimated as 40.37 (31.47) % for near (far) data. The same detection efficiency in [2] is used in this analysis.

*3.2. Remaining background*
There are still some background remained in the IBD candidate events passing the selection criteria. The methods to estimate the remaining background are well described in [2] and using the same method for the 1500 live days of RENO data we estimated the remaining background and summarized in Table 1.

**Table 1.** Event rates (per day) of the observed IBD candidates and the estimated background in 1.2 < $E_p$ < 8 MeV.

| Detector | Near | Far |
|---|---|---|
| Selected candidate events | 732,168 | 68,055 |
| Total background rate | 9.34±0.37 | 1.95±0.15 |
| IBD rate after background subtraction | 463.80±0.66 | 46.75±0.24 |
| Live time (days) | 1547.35 | 1397.78 |
| Accidental rate | 2.07±0.02 | 0.38±0.01 |
| $^9$Li/$^8$He rate | 5.49±0.36 | 0.93±0.15 |
| Fast neutron rate | 1.74±0.02 | 0.35±0.01 |
| $^{252}$Cf rate | 0.04±0.01 | 0.28±0.02 |

**4. Systematic Uncertainties**
The systematic uncertainties are obtained using the same methods described in [2]. Using these methods we estimated our systematic uncertainties on background and summarized in Table 2. The systematic uncertainties of reactor, detection efficiency including timing veto, and energy scale remain the same as before [2].

**Table 2.** Background systematic uncertainties in $1.2 < E_p < 8$ MeV.

|  | Bin-correlated | Bin-uncorrelated |
|---|---|---|
| Total background | 0.60 % (near), 1.99 % (far) | 3.94 % (near), 2.71 % (far) |
| Accidental | 0.37 % (near), 0.96 % (far) | 0.18 % (near), 0.49 % (far) |
| $^9$Li/$^8$He | 1.01 % (near), 3.66 % (far) | 6.71 % (near), 4.17 % (far) |
| Fast neutron | 0.23 % (near), 0.54 % (far) | 0.75 % (near), 0.83 % (far) |
| $^{252}$Cf | 6.00 % (near), 1.11 % (far) | 10.23 % (near), 12.62 % (far) |

## 5. Results

Using 1500 live days of data we updated the 5 MeV excess observation and spectral measurement of $\sin^2 2\theta_{13}$ and $|\Delta m^2_{ee}|$. Each study is separately described below.

### 5.1. 5 MeV Excess

The 5 MeV excess was first reported in 2014 using 800 live days of RENO data [3] where the correlation between the 5 MeV excess and the IBD rate, i.e. the reactor thermal power was also reported. These results are updated using 1500 live days of data. Figure 1 top panels show observed IBD prompt spectra of near and far data compared to the expected ones by the Huber and Mueller model [4,5] normalized to the area except the 5 MeV excess region. The bottom panels of the Fig. 1 show the difference between the two spectra in the corresponding upper panels, where yellow bands represent uncertainties in the model. Both near and far data have clear excess around the 5 MeV. This excess amounts to ~2.5 % of total observed IBD events below 8 MeV.

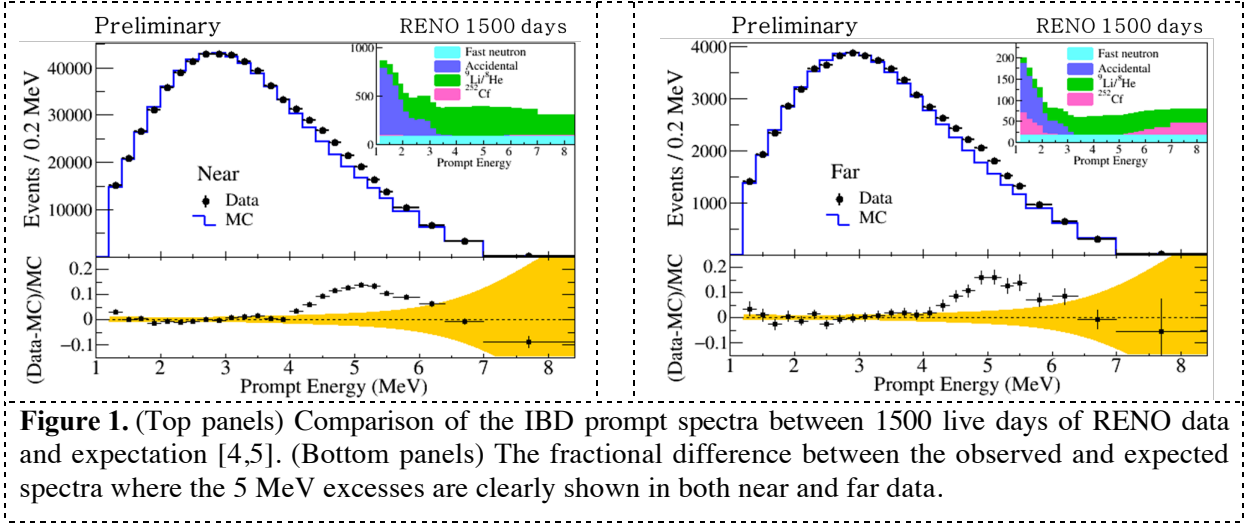

**Figure 1.** (Top panels) Comparison of the IBD prompt spectra between 1500 live days of RENO data and expectation [4,5]. (Bottom panels) The fractional difference between the observed and expected spectra where the 5 MeV excesses are clearly shown in both near and far data.

Figure 2 left panel shows a clear correlation between the 5 MeV excess and the reactor thermal power. We also observed a very weak (p-value = 0.24) correlation of the 5 MeV excess to the $^{235}$U fission fraction as shown in the right panel of Fig. 2, and reducing the uncertainty is necessary to draw any conclusion. Note that for these correlation studies 1800 live days of RENO near data was used to increase event statistics.

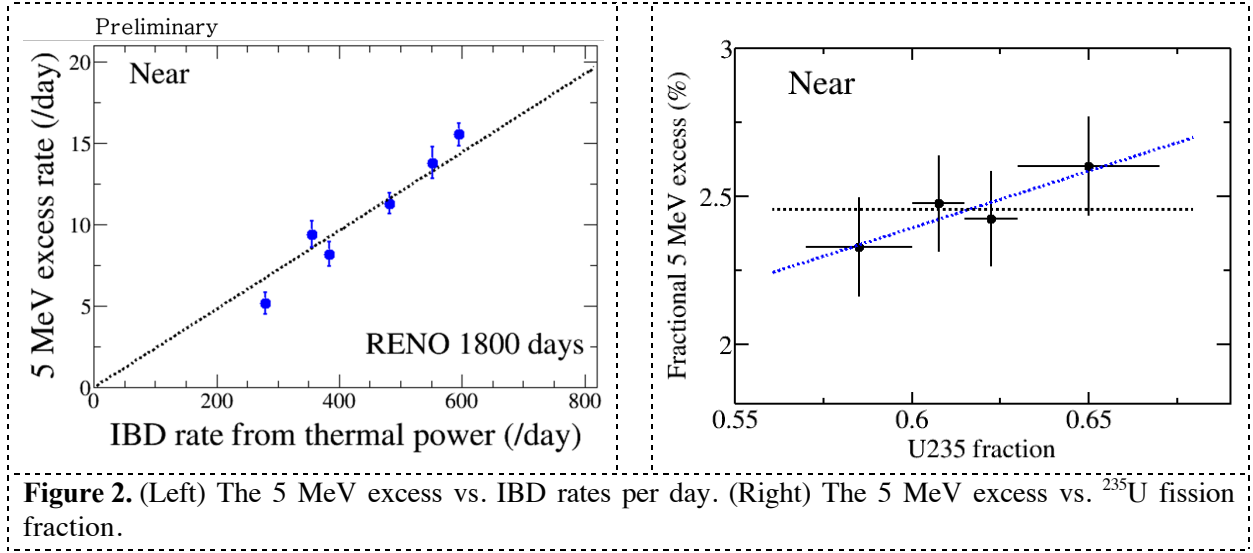

**Figure 2.** (Left) The 5 MeV excess vs. IBD rates per day. (Right) The 5 MeV excess vs. $^{235}$U fission fraction.

*5.2. Spectral measurement of $\sin^2 2\theta_{13}$ and $|\Delta m^2_{ee}|$*

Using the $\chi^2$ function for the rate + shape analysis described in [2], $\sin^2 2\theta_{13}$ and $|\Delta m^2_{ee}|$ are obtained for the 1500 live days of RENO data. The measured values using events in $1.2 < E_p < 8$ MeV are:

$$\sin^2 2\theta_{13} = 0.086 \pm 0.006 \text{ (stat.)} \pm 0.005 \text{ (syst.)}$$
$$|\Delta m^2_{ee}| = 2.61^{+0.15}_{-0.16} \text{ (stat.)} \pm 0.09 \text{ (syst.) } (\times 10^{-3} \text{ eV}^2).$$

The total uncertainty on $\sin^2 2\theta_{13}$ ($|\Delta m^2_{ee}|$) is reduced from 12 (10)% to 9 (7)% compared to our previous measurements using 500 live days of data [2]. Figure 3 top panel shows the observed IBD prompt spectrum at far (black dots with error bars) and the expected one obtained from near data assuming no oscillation. There is a clear discrepancy between the two due to electron anti neutrino disappearance at far, and their ratio is drawn in the bottom panel where the energy dependent discrepancy is shown well. Figure 4 shows the contour plot and the best-fit values of rate + shape (black dot) and rate-only (cross sign) measurements. Figure 5 shows the electron anti neutrino survival probability as a function of $L_{eff}/E$. Both near (open circles) and far (black dots) data points are shown with the best-fit oscillation probability (blue curve). The far data points matches very well to the best-fit oscillation. The near data points, however, matches extremely well to the best-fit oscillation since the near expectation without oscillation was obtained by unavoidably using near data itself rather than MC. Note that MC can not be used in this case because of the mismatch in the 5 MeV excess region.

## 6. Summary and Prospects

Using 1500 live days of data RENO has reduced the uncertainties to 9 % and 7 % for the $\sin^2 2\theta_{13}$ and $|\Delta m^2_{ee}|$ measurements, respectively. RENO expects to reduce the $\sin^2 2\theta_{13}$ uncertainty to ~6 % using data taken by 2018. With additional 2 or 3 more years of data taking from 2019 the uncertainty on the $|\Delta m^2_{ee}|$ measurement is expected to be reduced to 4~5 % even though the $\sin^2 2\theta_{13}$ uncertainty would remain as ~6 %.

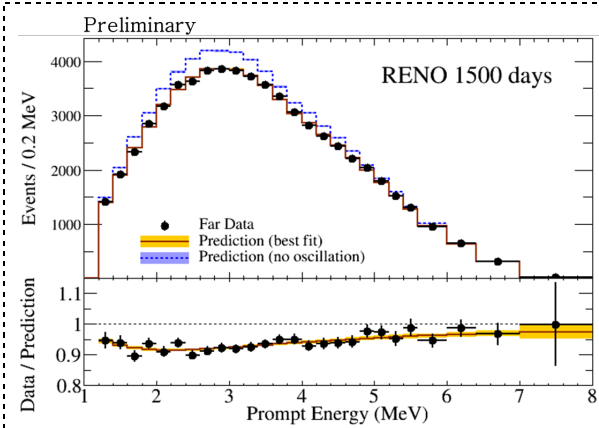

**Figure 3.** (Top panel) Observed (black dots with error bars) vs. expected (blue dotted histogram) IBD prompt energy spectra after background subtraction at far site. The expected spectrum at far site is obtained using the near IBD data assuming no oscillation. The orange histogram represents the expected IBD spectrum with best-fit oscillation parameters. (Bottom panel) Ratio of the observed to the expected IBD prompt spectra. There is a clear energy dependent reactor neutrino disappearance.

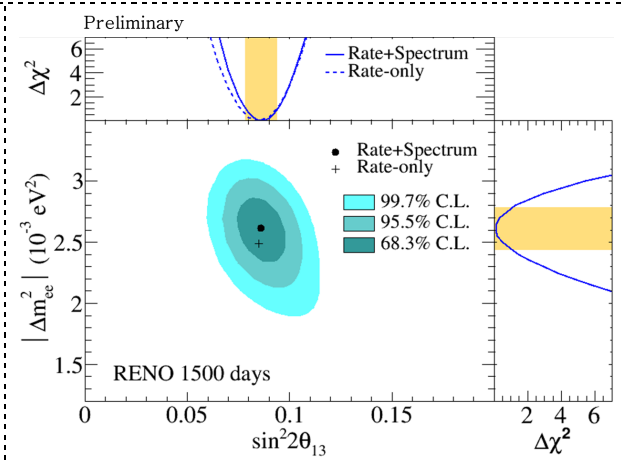

**Figure 4.** Contour plot of $\sin^2 2\theta_{13}$ vs. $|\Delta m^2_{ee}|$. The best fit value for rate + shape (rate-only assuming $|\Delta m^2_{ee}| = 2.49 \times 10^{-3}$ eV$^2$) analysis is represented as a black dot (cross). The three ellipses represent the corresponding confidence levels of 68.3%, 95.5%, and 99.7%. The upper (righter) panel shows 1-dimentional $\Delta\chi^2$ distribution for $\sin^2 2\theta_{13}$ ($|\Delta m^2_{ee}|$) and 1 σ error band in orange color.

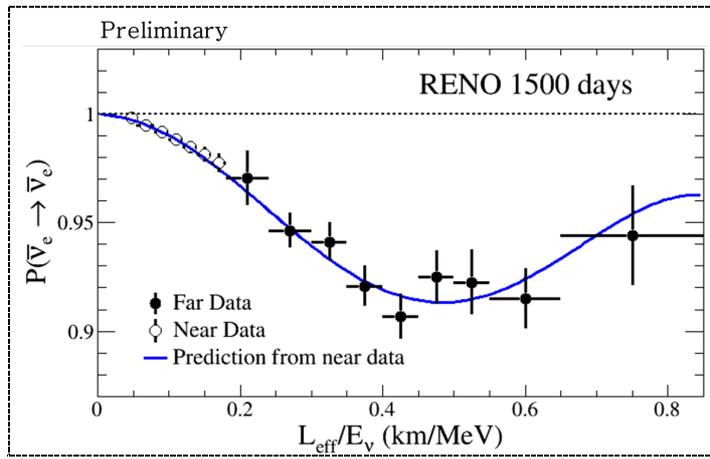

**Figure 5.** Reactor neutrino survival probability as a function of $L_{eff}/E$. The $L_{eff}$ is a flux weighted effective distance to a detector from the six reactors with different baselines. Far data matches well with the best-fit oscillation prediction (blue curve).